\documentclass[12pt,preprint]{aastex}

\shorttitle{McNeil's Nebula Object}
\shortauthors{Andrews, Rothberg, \& Simon}

\received{2004 April 8}
\begin{document}

\title{Mid-Infrared and Submillimeter Observations of \\ the Illuminating 
Source of McNeil's Variable Nebula}

\author{Sean M. Andrews, Barry Rothberg, and Theodore Simon}

\affil{Institute for Astronomy, University of Hawaii, 2680 Woodlawn Drive, 
Honolulu, HI 96822}
\email{andrews@ifa.hawaii.edu, rothberg@ifa.hawaii.edu, simon@ifa.hawaii.edu}

\begin{abstract}
We present post-outburst observations of the mid-infrared spectrum and 
submillimeter continuum of the illuminating source of the newly-discovered 
McNeil's Nebula in the L1630 region of Orion.  The 12\,$\mu$m flux of this 
source has increased by a factor of $\sim$25 after the outburst, whereas the 
submillimeter continuum remains at its pre-outburst level.  The bolometric 
luminosity has increased by at least an order of magnitude, to 
$\sim$34\,L$_{\odot}$, and is likely less than 90\,L$_{\odot}$.  The 
mid-infrared spectrum exhibits a strong and red continuum with no emission or 
absorption features.  The infrared slope of the spectral energy distribution 
characterizes the illuminating source as a flat-spectrum protostar, in both its 
active and quiescent states.  New CO spectral line observations show no 
evidence for a molecular outflow.  
\end{abstract}

\section{Introduction}
Recently, \citet{mcneil04} has reported the discovery of a bright reflection 
nebula in a previously dark region of the L1630 cloud within the Orion B 
complex.  An infrared source at the apex of McNeil's Nebula (hereinafter 
referred to as McNeil's Nebular Object, or MNO) has significantly brightened in 
the optical and near-infrared since 2003 November.  Detailed imaging and 
spectroscopy of that object are described by \citet{reipurth04}, 
\citet{briceno04}, and \citet{vacca04}.  Apparently the nebula has brightened 
in the past, indicating that MNO has resumed its outburst activity after 
roughly 37 years of quiescence \citep[see the discussions 
by][]{reipurth04,briceno04}.  

In the low state, MNO is positionally coincident with a faint $I$-band point 
source \citep{eisloffel97}, a 2MASS source (J05461313$-$0006058), an IRAS point 
source (05436$-$0007), an ISO source \citep{abraham04}, as well as a clump 
of submillimeter emission known as LMZ\,12 (Lis, Menten, \& Zykla 1999) and 
OriBsmm\,55 \citep{mitchell01}.  Previous work has shown that the L1630 cloud 
contains a number of deeply embedded young stellar objects (YSOs), which drive 
powerful outflows \citep[see][and references therein]{simon04}.  In the 
immediate vicinity of MNO there are several Class 0 (HH\,24-MMS, HH\,25-MMS) 
and Class I (SSV\,59, SSV\,63, HH\,26-IR) sources, as well as a series of 
outflows (HH\,22$-$27).  Both before \citep{eisloffel97,lis99} and after 
\citep{reipurth04} the current episode of activity, it has been suggested that 
MNO is the origin of the HH\,23 jet.  In this paper, we provide new 
mid-infrared and submillimeter observations of MNO in its high state and we 
discuss their implications for understanding the evolutionary status of that 
object. 

\section{Observations and Data Reduction}
Submillimeter continuum photometry at 450 and 850\,$\mu$m of MNO was obtained 
on UT 2004 March 10 with the Submillimeter Common User Bolometer Array (SCUBA) 
on the James Clerk Maxwell Telescope (JCMT) at Mauna Kea, Hawaii.  The 
precipitable water vapor (PPV) level was steady at $\sim$1.8\,mm, with frequent 
skydip observations showing zenith optical depths of 0.36 at 850\,$\mu$m and 
2.05 at 450\,$\mu$m.  The data were acquired in 20 sets of 18\,s integrations 
in a small 9-point jiggle pattern with the secondary mirror chopping 60\arcsec\ 
in azimuth at 7.8\,Hz.  Mars and CRL\,618 were observed to check the telescope 
pointing and to determine the absolute flux calibration.  The pointing has an 
rms error of $<$2\arcsec; the flux calibration is accurate to $\sim$10\% at 
850\,$\mu$m and $\sim$30\% at 450\,$\mu$m.  Standard reduction techniques were 
performed using the SURF software provided by the Joint Astronomy Center. 

Mid-infrared spectroscopic observations of MNO in the $N$-band were carried out 
on UT 2004 March 11 at the United Kingdom Infrared Telescope (UKIRT) on Mauna 
Kea with the Michelle imaging spectrometer  \citep{glass97}.  The Michelle 
detector is a Si:As 320$\times$240 pixel array operating between 8 and 
25\,$\mu$m.  In spectroscopy mode, the instrument uses a 91\arcsec\ slit (in 
our case with a P.A. = 0\degr\ orientation).  The Low-N grating and a 2-pixel 
wide (0\farcs76) slit were used to provide spectral coverage between 7.5 and 
12.5\,$\mu$m at a resolving power of $R\sim200$.  Atmospheric variations were 
removed using a standard nod-and-chop pattern during the total on-source 
integration time of $\sim$1250\,s.  The atmosphere was dry (PPV $\sim$1.6\,mm) 
and stable, providing FWHM values for point sources of 0\farcs8 at 10\,$\mu$m.  
Slit images demonstrate that MNO is point-like in the mid-infrared.

The spectra were sky-subtracted, flat-fielded and combined using the
Observatory Reduction and Acquisition Control (ORAC) software 
\citep{alasdair02}.  The IRAF task APALL was then used to extract spectra with 
a 3 pixel (1\farcs14) aperture.  A wavelength calibration solution was 
determined from terrestrial emission lines identified with ATRANS atmospheric 
models \citep{lord92}.  The rms of the wavelength fit was $\sim$0.04\,$\mu$m.  
The primary flux standard Rigel (BS\,1713) was used for both telluric 
correction and flux calibration.  The MNO spectrum was divided by the standard 
star spectrum and the result was scaled by a model thermal spectrum based on 
the $N$-band (10\,$\mu$m) flux of the standard star.  

Spectral line maps of $^{12}$CO\,(J=$2-1$) and $^{13}$CO\,(J=$2-1$) around MNO 
(covering 5\arcmin$\times$5\arcmin\ and 2\arcmin$\times$2\arcmin, respectively) 
were obtained from Mauna Kea with the 230-GHz heterodyne receiver on the CSO 
telescope on UT 2004 March 17.  The observations were conducted in the 
on-the-fly mode, providing Nyquist-sampled maps with a spatial resolution of 
$\sim$31\arcsec.  The 50-MHz AOS backend provided a spectral resolution of 
$\sim$0.06\,km\,s$^{-1}$ per channel and rms noise levels of $\sim$1.1\,K in 
both lines.  Standard procedures with the CLASS software package were used to 
reduce the data.  

\section{Results}
The broadband spectral energy distribution (SED) of MNO is shown in Figure 1 
for both the high and low activity states.  In the active state, the $I$-band 
flux is from \citet[][UT 2004 February 26]{briceno04}, the additional optical 
($g^{\prime}r^{\prime}i^{\prime}$) and near-infrared ($JHK^{\prime}$) data are 
from \citet{reipurth04}, and the $L$ and $M$-band fluxes are from the spectrum 
of \citet{vacca04}.  In the quiescent state, the data are the same as those 
used by \citet{abraham04} to analyze the pre-outburst SED, with the inclusion 
of an $I$-band flux point from \citet[][UT 1999 January 9]{briceno04}.  The 
standard photometric zero-points of \citet{bessell79}, \citet{fukugita96}, and 
\citet{tokunaga00} were used to convert magnitudes into flux densities.  The 
error bars in Figure 1 represent the systematic uncertainties in absolute flux 
calibration.  

Consistent with the factor of $\sim$16 near-infrared brightening of MNO, the  
12\,$\mu$m flux density rose by a factor of $\sim$25 during the high state.  In 
the standard classification scheme for young stellar objects, or YSOs 
\citep[][ Andr\'{e} et al. 2000]{lada84, adams87}, a SED type is assigned from 
the logarithmic slope of the spectrum, $n$, where $\nu F_{\nu} \propto 
\nu^{n}$.  For the conventional infrared indices of the post-outburst SED of 
MNO, we obtain $n=-0.62\pm0.14$ between 2 and 10\,$\mu$m (corresponding to the 
$K-N$ color) and $n=-0.66\pm0.13$ between 2 and 12\,$\mu$m\footnote{These 
indices are calculated from flux measurements taken $\sim$5 weeks apart, and 
are therefore subject to any changes in the near-infrared fluxes during that 
time.}.  The 2MASS 2\,$\mu$m and IRAS 12\,$\mu$m points from the quiescent 
phase of MNO yield an index of $n=-0.35\pm0.18$.  The foregoing values of the 
SED slope in the infrared establish MNO as a flat-spectrum/Class I object 
regardless of its activity level.  

Figure 2 shows the mid-infrared spectrum of MNO from 7.5 to 12.5\,$\mu$m.  The 
spectrum exhibits a bright, red, featureless continuum across the $N$-band.  A 
simple power-law fit to the continuous spectrum gives $F_{\nu} \propto 
\lambda^{2.5}$, which corresponds to $n=-1.5 \pm 0.1$.  The positions of 
various solid-state features that are commonly seen in the spectra of other 
YSOs are labeled in Figure 2, although none of them are detected here.  

The flux densities at 450 and 850\,$\mu$m measured in our SCUBA observations of 
MNO are $1.589\pm0.099$ and $0.316\pm0.005$\,Jy\,beam$^{-1}$ (quoted errors are 
1$\sigma$ statistical uncertainties), and correspond to FWHM apertures of 
$\sim$9\arcsec\ and $\sim$14\arcsec, respectively.  Our 850\,$\mu$m flux 
density is essentially identical to the value cited by \citet{mitchell01} for a 
$\sim$20\arcsec\ diameter aperture, based on SCUBA maps of L1630 made by those 
authors in 1998.  \citet{lis99} have argued that the 1.3\,mm emission is 
confined within a deconvolved diameter of $<3$\arcsec.  Given the agreement 
between the 850\,$\mu$m flux densities measured in two different sized 
apertures, the circumstellar structure responsible for the submillimeter 
emission must also be compact, and $\lesssim10^4$\,AU in diameter if we assume 
the accepted distance of $d\sim450$\,pc for L1630.  

Contrary to the behavior of the optical and infrared spectrum of MNO, no 
changes are apparent in the submillimeter continuum (350\,$\mu$m $\lesssim 
\lambda \lesssim$ 1.3\,mm) as a result of the current activity.  Assuming the 
Rayleigh-Jeans limit and optically-thin emission, the submillimeter flux 
density follows the relationship $F_{\nu} \propto \nu^{2+\beta}$, where $\beta$ 
is the power-law index of the grain opacity.  A least-squares fit to the four 
submillimeter points in the ($\log{\nu},\log{F_{\nu}}$) plane gives 
$\beta=0.65\pm0.19$.  The slightly more sophisticated model of \citet[][their 
Equation 2]{mitchell01}, if applied to the 450 and 850\,$\mu$m measurements, 
yields a range of $\beta$ between 0.79 and 1.24 (for dust temperature values 
$T_{dust} \sim 50$ and 20\,K, respectively).  The discrepancy in $\beta$ 
between a simple fit and the \citet{mitchell01} model may indicate that the 
material is not optically-thin and/or the Rayleigh-Jeans approximation is 
inappropriate.  The value of $\beta$ is apparently near unity, although we 
caution that it would be best determined with a full SED model fit if more 
far-infrared data become available while MNO is in an active state.

\section{Discussion}
As shown above and discussed by \citet{abraham04}, the small (slightly 
negative) infrared spectral index, $n$, establishes MNO as a 
flat-spectrum/Class I object.  The large spectral slope in the submillimeter 
is also more typical of Class I sources than the presumably more evolved Class 
II sources \citep{dent98}.  It is worthwhile to compare MNO with the other 
young outflow sources in L1630, since it is reasonable to assume these other 
objects, as siblings of MNO, may experience similar outbursts.  Using our data 
and the flux values given by \citet{mitchell01}, we note that the 850\,$\mu$m 
flux decreases by a factor of 4 between the Class 0 (e.g., HH 24-MMS and
HH 25-MMS) and Class I (e.g., SSV 59 and SSV 63EW) objects in 
L1630, and by another factor of 4 from the Class I objects to MNO.  With an 
appropriate correction for distance, flat-spectrum sources in other clouds have 
850\,$\mu$m fluxes similar to that of MNO (e.g., HL Tau).  Applying the 
standard assumption of optically-thin, isothermal dust \citep[$\kappa_{850} = 
0.03$\,cm$^{2}$ g$^{-1}$, $T_{dust} = 50$\,K;][]{beckwith99}, we infer from
the 850\,$\mu$m flux a total (gas + dust) circumstellar mass of $M_{cs} \sim 
0.06$\,M$_{\odot}$.  Our mass estimate is an order of magnitude lower than that 
calculated by \citet{abraham04}, primarily due to the different $T_{dust}$ 
assumed.  Nevertheless, $M_{cs}$ is significantly larger than the values 
typically found for Class II disks.  The submillimeter evidence therefore 
agrees well with the interpretation from the infrared SED that MNO is currently 
in transition from the Class I to Class II stage of protostellar evolution.  
The circumstellar environment of MNO is probably dominated by a massive 
accretion disk, with only a remnant envelope structure.

The mid-infrared spectrum of MNO shown in Figure 2 is notably devoid of any 
solid-state spectral features.  None of the three major PAH features (at 7.7, 
8.6, and 11.2\,$\mu$m) are seen, in agreement with the absence of the more 
frequently detected 3.3\,$\mu$m PAH feature in the 1--5\,$\mu$m spectrum 
presented by \citet{vacca04}.  PAH features are seldom observed in low-mass 
YSOs (Smith, Sellgren, \& Tokunaga 1989; Brooke, Sellgren, \& Geballe 1999), 
but are more common in their higher mass counterparts, the HAeBe stars 
\citep{ressler03,habart04}.  There is no evidence for 3.3\,$\mu$m PAH emission 
in the infrared spectra of the other Class I objects in L1630 discussed by 
\citet{simon04}.  More puzzling is the absence of a 9.7\,$\mu$m amorphous 
silicate feature in the spectrum of MNO.  According to \citet{cohen84}, at 
least two of the other Class I objects in L1630 (SSV\,59 and 63) have strong 
silicate absorption bands.  The optical depth of the 3.1\,$\mu$m H$_2$O ice 
band measured by \citet{vacca04}, $\tau_{ice} \sim 0.7$, suggests that the 
9.7\,$\mu$m feature of MNO should also be in absorption.  Using information 
provided by \citet[][their Figure 8]{brooke99}, we would expect a band depth of 
$\tau$(9.7\,$\mu$m) $\sim0.4$.  Such a shallow absorption could be entirely 
filled in by emission from the optically-thin surface layers of a disk if it is 
viewed at a suitably high inclination angle \citep[][their Figure 
5]{chiang99}.  

Recent VLA observations at wavelengths of 3.6, 6, and 20\,cm detected no radio 
emission from MNO to rms noise levels of 42, 48, and 140\,$\mu$Jy, respectively 
\citep{claussen04}.  The flux densities expected from dust, based on an 
extrapolation from the submillimeter wavelengths, fall below these detection 
limits.  Of the Class I sources in L1630, radio detections have been reported 
previously for SSV\,63 \citep[three separate components:][]{reipurth04b} and 
HH\,26-IR, whereas SSV\,59 has a 3$\sigma$ upper limit of 66\,$\mu$Jy 
\citep{gibb99}.  The radio emission from objects like SSV\,59 and MNO is either 
absent or optically-thick.  

The lack of an obvious high velocity jet or molecular outflow from MNO remains 
to be explained and further distinguishes this object from the other YSOs in 
L1630.  Although a powerful wind is evident in post-outburst optical and 
infrared spectra, shocked emission from the standard optical and infrared  
forbidden lines is absent \citep{reipurth04,briceno04,vacca04}.  The 
submillimeter CO spectral line maps obtained by us and by \citet{lis99} show no 
spatial distinction between redshifted and blueshifted emission.  The limited 
spatial resolution of these maps does not definitively rule out the presence of 
a molecular outflow, and so interferometric observations will be required to 
address this issue appropriately.  However, as shown in Figure 3, our 
single-dish observations do rule out the presence of gas moving at velocities 
greater than $\sim$3\,km s$^{-1}$ from the rest velocity of the L1630 cloud.  
The best candidate outflow signatures from MNO are the HH\,23 clumps of 
[\ion{S}{2}] emission, which are located $\sim$3\arcmin\ to the north, along 
the direction of McNeil's Nebula \citep{eisloffel97,lis99,reipurth04}.  

Using a simple trapezoidal integration of the SEDs shown in Figure 1, we 
estimate that the bolometric luminosity of MNO has changed from $\sim$3.5 to 
34\,L$_{\odot}$ during the outburst.  The pre-outburst value of $L_{bol}$ was
estimated by \citet{lis99} to be 2.7\,L$_{\odot}$ and by \citet{abraham04} 
to be 5.6\,L$_{\odot}$.  The former result is identical to ours  when allowance 
is made for the different distances that were used.  The \citet{abraham04} 
value is slightly larger than ours because those authors chose to correct 
for an assumed (non-local) extinction of $A_V = 13$\,mags.  However, 
such a correction leads to an over-estimate of $L_{bol}$ if the extinction 
is due to local circumstellar material (as is thought to be the case for MNO) 
because the short-wavelength flux is thermalized and then re-emitted at 
longer wavelengths.  Our post-outburst value of $L_{bol}$ likely underestimates 
the true value because we lack information in the SED near its peak in the 
far-infrared.  We have used a linear extrapolation between the 
infrared and submillimeter portions of the SED to estimate a peak flux of 
$\sim$80\,Jy (at roughly 70\,$\mu$m), which leads to an upper limit on the
 post-outburst $L_{bol} \le 90$\,L$_{\odot}$.   

\citet{briceno04} have calculated an intrinsic luminosity of the post-outburst 
MNO of $L = 219$\,L$_{\odot}$ from their $I$-band data, assuming that the 
spectral type is A0\,V and that $A_I = 7.2$, or $A_V \sim 15$.  However, the 
post-outburst colors of MNO suggest a lower extinction of $A_V \sim 11$ 
\citep{reipurth04}, which is in good agreement with the value indicated by the 
depth of the infrared H$_2$O ice band \citep{vacca04}.  Using the interstellar 
extinction law of \citet{mathis90},  the main sequence colors and bolometric 
corrections of \citet{kenyon95}, and the lower post-outburst extinction value, 
the \citet{briceno04} luminosity is reduced by more than a factor of 4, to 
$L_{bol} \sim 47$\,L$_{\odot}$ (this assumes $R_V = 3.1$; the value would be 
roughly a factor of 2 larger if $R_V = 5.0$).  The revised value is thus in 
good agreement with the estimate obtained by integrating under the SED.  Both 
analyses, all the same, should be treated with some caution.  In the 
post-outburst stage, there will be a considerable contamination of the $I$-band 
magnitudes from scattering in McNeil's Nebula (particularly with the large 
pixel scale in those observations).  Moreover, it is not at all clear that main 
sequence colors and bolometric corrections apply to YSOs like MNO.  At the same 
time, a simple integration of the SED includes any luminosity from accretion, 
and consequently may not be an accurate representation of the intrinsic 
photospheric flux (including the portion that is thermalized by local dust and 
re-emitted at longer wavelengths).  With those caveats in mind, the $L_{bol}$ 
values both before and after the outburst do not support earlier claims that 
MNO is a B type star, to order of magnitude they are similar to those of other 
YSOs in L1630 \citep{cohen84,berrilli89}, and they are significantly lower than 
those derived for FU Orionis stars \citep[e.g.,][their Table 1]{sandell01}.

\section{Summary}
We have presented new observations in the mid-infrared and submillimeter of the 
outburst star which illuminates McNeil's Nebula.  The object has brightened by 
a factor of $\sim$25 in the mid-infrared, yet remained at the same brightness 
in the submillimeter.  The bolometric luminosity has increased by (at least) an 
order of magnitude, but still remains low compared to that expected from either 
an early-type photosphere or a FUor.  Our limited CO spectral-line maps 
indicate that no high-velocity flows of molecular gas have yet appeared in the 
immediate vicinity of this object.  Far-infrared observations of MNO from the 
\emph{Spitzer} Space Telescope would help to complete the post-outburst SED and 
aid in determining $L_{bol}$ and $T_{dust}$, thus $M_{cs}$, providing important 
evolutionary constraints on this object.  Observations of solid-state features 
in the mid- and far-infrared spectrum of MNO would also serve to provide a 
constraint on the disk inclination angle and thus would offer a test of our 
explanation for the absence of the 9.7\,$\mu$m feature.

\acknowledgments We are grateful to Tom Kerr, Sandy Leggett, Thomas Lowe, 
Sandrine Bottinelli, and Jonathan Williams for their assistance with the 
observations and data reduction.  Suggestions from an anonymous referee have 
greatly improved this Letter.  We also want to thank Mark Claussen, Mike 
Cushing, Joel Kastner, Bo Reipurth, and Bill Vacca, who have generously shared 
their results before publication.

\clearpage
\begin{figure}
\plotone{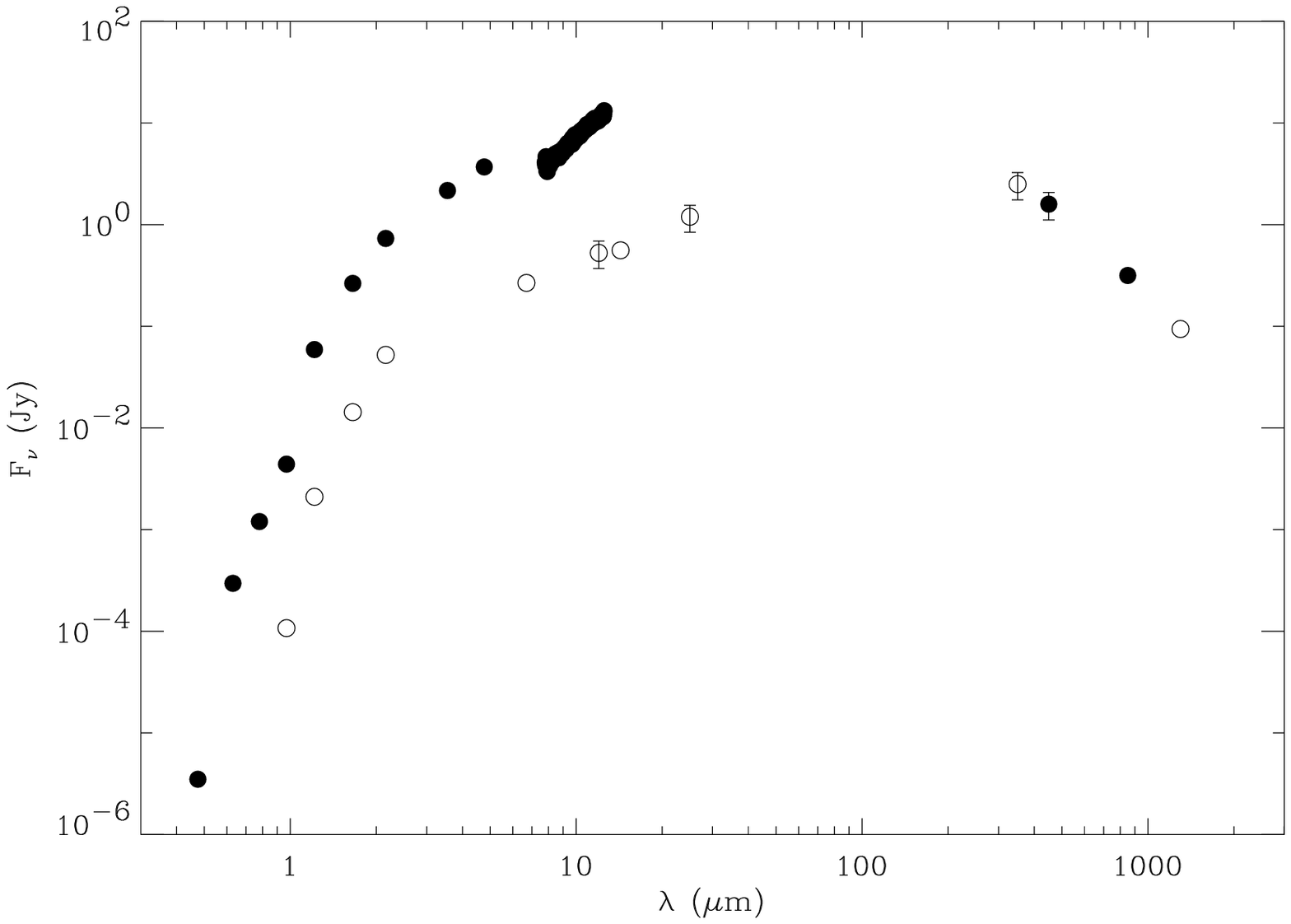}
\figcaption{The broadband SED of MNO.  Open circles mark data taken in the 
quiescent state before 2003 November, and filled circles mark post-outburst 
data from 2004 February and March (see the text for references).  The 
mid-infrared spectrum is  shown in greater detail in Figure 2.  The continuum 
slope in the infrared indicates that MNO is a flat-spectrum/Class I protostar.}
\end{figure}

\clearpage
\begin{figure}
\plotone{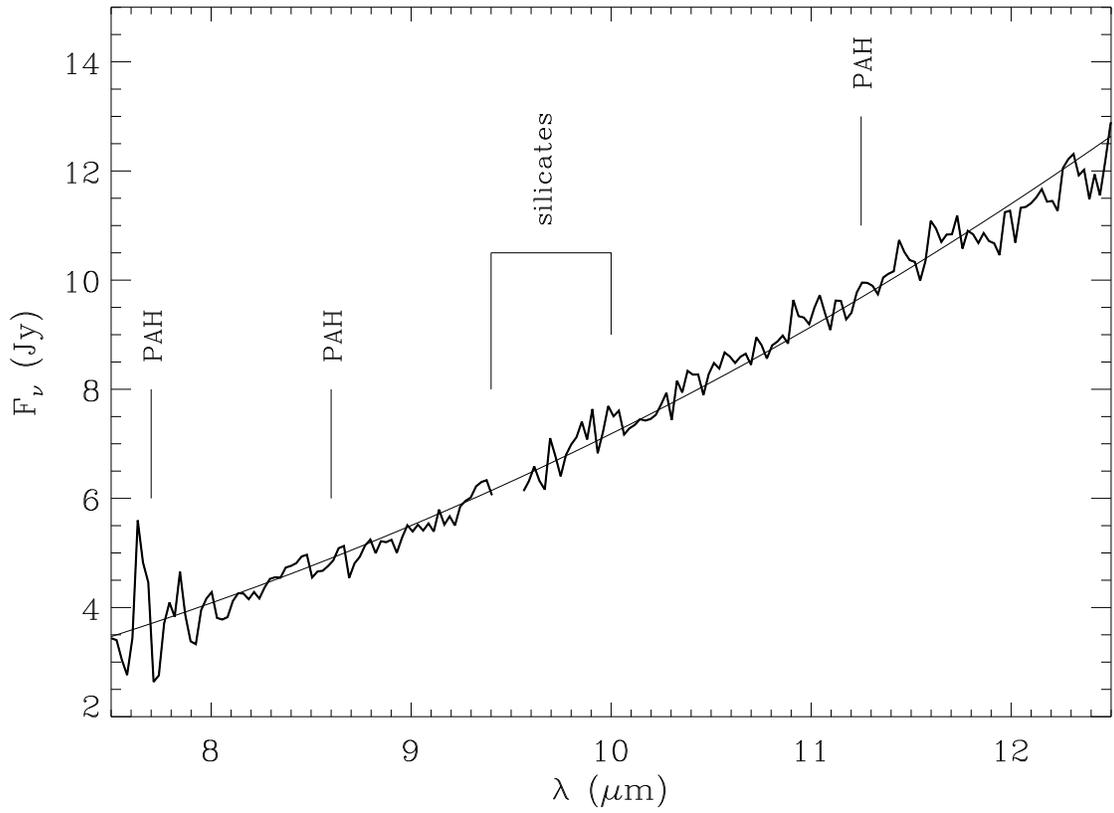}
\figcaption{The mid-infrared spectrum of MNO.  A 5-pixel gap near 9.4\,$\mu$m 
has been blanked out due to a residual telluric ozone line.  Although no 
spectral features have been detected, the positions of some of the more common 
YSO features are indicated.  A simple power-law fit to the continuum ($F_{\nu} 
\propto \lambda^{2.5}$) is shown for comparison.}
\end{figure}

\clearpage
\begin{figure}
\plotone{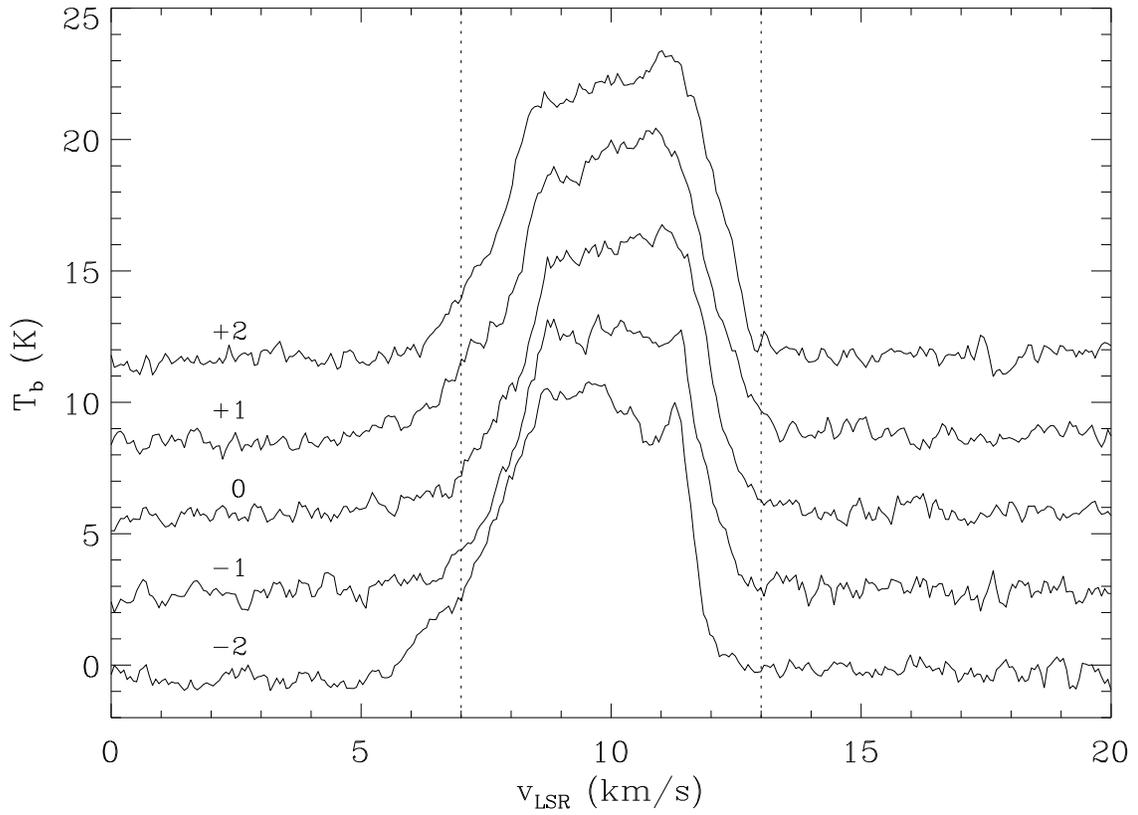}
\figcaption{Composite CO\,(J=2$-$1) spectra from the region surrounding MNO.  
Each spectrum is the average in a 1\arcmin\ diameter aperture, labeled 
according to its offset in declination in arcminutes from the position of MNO.  
The spectra are offset in $T_b$ by 3\,K for clarity.  The dotted lines mark 
velocity offsets of $\pm$3\,km s$^{-1}$ from the rest velocity of the L1630 
cloud.}
\end{figure}

\end{document}